\documentclass[oldversion]{aa}  
\usepackage{graphicx}
\usepackage{txfonts}
\def\bibl{\parindent=0pt \hangindent=0.150 in }
\begin{document}
   \title{Simulation of the interstellar scintillation and the extreme scattering events of pulsars}

   \subtitle{}

   \author{M. Hamidouche
          \inst{1,2}
          \and
          J.-F. Lestrade\inst{3}
          }

   \offprints{M. Hamidouche}

   \institute{LPCE/CNRS, 3A Ave de la Recherche Scientifique, F45071 - Orl\'eans, France\\                  
         \and
         Astronomy Department, University of Illinois at Urbana-Champaign,
              1002 W Green St, Urbana, IL 61801, USA\\
            \email{mhamidou@astro.uiuc.edu}\\
                                \and
             LERMA/CNRS, Observatoire de Paris, 77 Ave Denfert Rochereau, F75014 - Paris, France\\
             \email{jean-francois.lestrade@obspm.fr}\\
             }

   \date{Received/Accepted}

% \abstract{}{}{}{}{} 
% 5 {} token are mandatory
 
  \abstract{The rare and conspicuous flux density variations of some radio sources (extragalactic and pulsars) 
for periods of weeks to months have been denoted Extreme Scattering Events (ESE's) by Fiedler et al. (1987).
Presently, there is no astrophysical mechanism that satisfactorily produces
this phenomenon. In this paper, we conjecture that inhomogeneities of the
electronic density in the turbulent interstellar medium might be the origin of this phenomenon. We
have tested this conjecture by a simulation of the scintillation of the pulsar B1937+21 at 1.4 GHz and 1.7 GHz for a period of six months. To this end, 
we have constructed a large square Kolmogorov phase screen made of 
$131{\rm k} \times 131{\rm k}$ pixels with electron inhomogeneity scales 
ranging from  $6~10^{6} {\rm m}$ to $10^{12} {\rm m}$  
and used the Kirchhoff-Fresnel integral to simulate dynamic
spectra of a pulsar within the framework of Physical Optics. 

The simulated light curves exhibit a 10 day long variation 
simultaneously at 1.41 and 1.7~GHz that is alike
the ``ESE'' observed with the Nan\c cay radiotelescope toward the pulsar B1937+21 in October 1989.
Consequently, we conclude that ``ESE'' toward pulsars 
can be caused naturally by the turbulence in the
ionized interstellar medium instead of invoking the crossing of 
 discrete over pressured ionized clouds on the line of sight
as in the model of Fiedler et al. (1987). We suggest that longer 
 events could occur in a simulation of scintillation,
 if larger electron inhomogeneities $(> 10^{12} {\rm m})$ were
included in the construction of the Kolmogorov phase screen. This next step
requires a supercomputer.}

   \keywords{Scattering, Turbulence, ISM : structure, Stars : pulsars : B1937+21}

   \maketitle
%
%________________________________________________________________

\section{Introduction}\label{intro}

The rare and conspicuous flux density variations
of some radio sources  for periods of weeks to months 
have been denoted Extreme Scattering Events (ESE's).
Usually, the flux density drops by at least 50 \% and increases
somewhat at ingress and egress although there is a variety of possible 
shapes.  The first event was  recognized  
 in the direction of the quasar 0954+658 at 2.7 and 8.4 GHz
(Fiedler et al. 1987). The first event associated with
a pulsar was detected in the direction of the millisecond pulsar B1937+21 
at 1.41 GHz (Cognard et al. 1993).  
The most recent census indicates that 15 ESE's  have been identified
in the radio light curves of 12 quasars in the entire Green Bank 
Interferometer monitoring of 149 radio sources at GHz frequencies 
between 1979 and 1996 (Lazio et al. 2001). Five ESE's have been identified in the direction of the pulsar B1937+21 between 1989 and 1996
(Lestrade, Rickett \& Cognard 1998, LRC98 hereafter) and a very long
event was observed in the direction of the pulsar J1643-1224 (Maitia, 
Lestrade \& Cognard 2003). Hill et al. (2005) also report criss-cross
pattern in the dynamic spectra of PSR B0834+06 at 327MHz they interpret as 
resulting from compact scattering objects essentially stationary in the screen in 20 years.   
  
The mechanism usually invoked for this phenomenon
is scattering by a plasma lens that occurs when
a discrete ionized cloud crosses the line of sight.
Fiedler et al. (1987) proposed that the
medium inside the cloud is highly turbulent, causing extreme scattering 
responsible for the flux density variation and  source broadening. 
Differently, Romani et al. (1987) proposed that the medium inside the 
cloud acts as a purely refractive lens producing 
extreme refraction with caustics responsible for the flux variation and
source displacement during the event. 
The observations of ESE's in directions of quasars and pulsars  
and these models provide an estimation of the cloud size of 1-50~AU and internal electronic 
density of $100-1000~{\rm e^-}~{\rm cm^{-3}}$ (Fiedler et al. 1994). Therefore, such a structure
 is over pressurized relative to the ambient 
medium making its lifetime as short as  a few years. In addition, the observed 
rate of occurrence implies a cloud space density as high as 
$\sim 10^6$~clouds~pc$^{-3}$ (LRC98). 
There is no known mechanism that can produce
so many discrete ionized  clouds in the ISM. In order to 
avoid this difficulty, we conjecture in this paper that  
the Kolmogorov turbulence of the ionized
ISM is the  natural cause for the ESE's and their occurrence 
is statistical. 

In order to support this proposition, we have simulated  
interstellar scintillation at 1.41 and 1.7 GHz toward the pulsar B1937+21 to compare to the observations taken
 at the Nan\c cay radiotelescope\footnote{Nan\c cay radiotelescope is an instrument of the Observatoire de Nan\c cay, France}. 
The flux density of this pulsar measured at the Nan\c cay radiotelescope
is an average over an integration time of 70 minutes and
an integration bandwidth of 7.5 MHz centered around 1.41 GHz and 12.5 MHz around 1.7 GHz.  These intervals 
are large enough to average several diffractive interstellar scintillation patterns in the time-frequency domain, 
and so essentially  sample the refractive regime. However, some significant influence of diffractive ISS remains. Therefore, we
conducted a full scintillation calculation including the diffractive and refractive scales.
The refractive time scales measured at Nan\c cay in the direction of  B1937+21
 are 16$\pm$1 days and 8$\pm$1 days, at 1.41 GHz and 1.7 GHz respectively.
Also, the modulation indexes are $m_r$ = 0.30$\pm$0.02 and 0.50$\pm$0.04 at 1.41 GHz and 1.7 GHz respectively. The zero-lag 
correlation of the flux density between these two frequencies is 0.93$\pm$0.05. Detailed descriptions 
of these measurements are given in LRC98.

For the simulation presented in this paper, 
we have used the Kolmogorov 3-D power spectrum of the electronic density $\delta {n_e(x,y,z)}$ in the interstellar medium as established locally ($< 1 $kpc) by Armstrong, Rickett \& Spangler (1995). We have used the Kirchhoff-Fresnel integral to compute pulsar dynamic spectra in the framework of Physical Optics.
In section 2, we present the formulae for this simulation carried out 
in the thin screen approximation, the construction of the Kolmogorov
phase screen and the computation of the diffraction pattern (see also Hamidouche 2003). In section 3, we construct a large square Kolmogorov phase screen
of $131{\rm k} \times 131{\rm k}$ pixels.
In section 4, we make a series of tests to empirically 
determine how to set the parameters of the Kirchoff-Fresnel integral. 
In section 5, we present the result of the simulation in the direction of B1937+21 simulated over a period of 6 months.  

\section{Scattering of a Point-Like Source by the Ionized Interstellar Medium}\label{iss} 

The historical problem of diffraction have been cast in modern mathematical
terms by Born and Wolf (1999) and Goodman (1968). A particularly 
clear presentation of this problem applied to interstellar optics 
in the radio domain is in Narayan (1992) and 
Gwinn et al. (1998). The tenuous
interstellar plasma is a random medium where inhomogeneities 
of the density of free electrons produce fluctuations of the 
index of refraction. The radio wavefronts emitted by  a 
point-like source (pulsar) and propagating through  such a medium 
produce rapid intensity fluctuations  at the Earth.
These fluctuations are due to
diffraction speckles in the plane of the observer with 
a short time scale of several minutes and over a narrow bandwidth ($\sim 1$ MHz) at radio frequencies.
This is the DISS phenomenon (Diffractive interstellar medium scintillation).
This random medium produces also slow intensity variations of 
 10-100~\% over several days or more  
and  over a broad bandwidth ($>$100 MHz). 
This is the RISS phenomenon (Refractive interstellar medium scintillation).
 Rickett (1988) proposes  a  concise
presentation of these phenomena and of their  main observables. 
Rickett (1977), Blandford \& Narayan (1985) and Rickett (1990) 
provide a complete  derivation of these
observables as functions of the random medium physical properties. 
Diffractive scintillation in dynamic spectra
of pulsars is caused by  medium inhomogeneities, whose scale $s_0$  
is  called  diffractive scale or coherence length.
From observations of pulsars
at frequency around 1~GHz, this diffractive scale is 
between $10^5$~m and $10^8$~m. Refractive scintillation
 of pulsars depends on
medium inhomogeneities whose  scales  are about the scattering disk radius
 $ r_S =  \theta_S L$ where $L$ is the distance to the 
equivalent turbulent screen and $\theta_S = {\lambda \over {2 \pi s_0}}$
is the scattering angle. 
From pulsar observations, this refractive scale $r_S$ is between 
$10^{10}$~m and $10^{13}$~m. 
These two main scales of the ionized interstellar medium are suggestively
sketched in  Cordes, Pidwervetsky \& Lovelace (1986).
The line of sight to a pulsar moves through the medium with a
significant transverse velocity  $ V_{\bot}$  and continuously samples
inhomogeneities of sizes $s_0$  and $r_S$ 
so that the diffractive  timescale is
 $ t_d = s_0 / V_{\bot}$ and the refractive timescale 
is $ t_r = L \theta_S / V_{\bot}$. The decorrelation
bandwidth $\Delta \nu_{dc} = 1 / 2 \pi \tau_S$ 
observed in dynamic spectra of pulsars
is associated with the temporal broadening of pulsars $\tau_S \sim  L \theta_s^2 /2c$ (Rickett 1988). The refractive 
modulation index ($m_r$ = rms of intensity / mean intensity)
is an important quantity directly comparable to observations
and is theoretically  
$m_r = 1.08 (\Delta \nu_d /  \nu)^{1/6}$ (Appendix B in Gupta, Rickett
\& Coles 1993). Pulsar dynamic spectra provide  measurements of  
the diffractive parameters 
$t_d$  and $\Delta \nu_{dc}$ that  yield the turbulence strength $C_n^2$ 
({\it e.g.} Gupta, Rickett \& Lyne 1994).  Long term pulsar flux density series yields the refractive parameter $t_r$ that is typically 
in tens of days  and the modulation index $m_r < 1 $ (Stinebring et al. 2000).

Armstrong, Rickett, \& Spangler (1995) analyze a large {\sl corpus} 
of radioastronomy observations to establish that the 3-D space
power spectrum ${P_{3N}}(q_x,q_y,q_z)$ of the free electron density  
$\delta { n_e} (x,y,z)$
in the local interstellar medium ($ < 1 $ kpc)  
is of the Kolmogorov turbulence type. They give evidence of the power spectrum 
 ${{P_{3N} (q_x,q_y,q_z)} = C_n^2} \times \left(\sqrt{q_x^2+q_y^2+q_z^2}\right)^{-11/3}$ valid  over 6 decades, $10^{-12}\;{\rm m^{-1}}< \sqrt{q_x^2+q_y^2+q_z^2} < 10^{-6}\;{\rm m^{-1}}$. They suggest that it might extend over 12 decades.  The derived turbulence strength by the same authors is $C_n^2~\sim~10^{-3}~{\rm m}^{-20/3}$. This value can be considered as typical but large deviations are found depending on the celestial direction $ 10^{-4}~<~C_n^2~<~10^{-1}~{\rm m}^{-20/3}$ (Johnston, Nicastro \& Koribalski 1998; Cordes, Weisberg \& Boriakoff 1985).
   
The duration of ESE's (weeks to years; Maitia, Lestrade \& Cognard 2003) 
and the typical
transverse velocity of pulsar (tens of km/s; Gupta, Rickett \& Lyne 1994)
would make these events
associated with large-scale  inhomogeneities of free electrons,
{\it i.e.}  the refractive scales $r_S$ ($10^{10}$~m and $10^{13}$~m) following
our conjecture.
To test this concept, we simulate the scintillation 
of the pulsar B1937+21 at 1.41 and 1.7 GHz with a Kolmogorov phase screen
as large as $10^{12} {\rm m} \times 10^{12} {\rm m}$ sampled by
 $131{\rm k} \times 131{\rm k}$ pixels.
The 3-D nature of this simulation is reduced advantageously 
to a 2-D problem in the thin screen approximation valid in the interplanetary
and interstellar tenuous plasma. Following Salpeter (1967) and Lovelace (1970),
this approximation is the limit where the ray lateral deflection is weak relative to the radial 
phase fluctuations. In this condition, the phase of a wavefront reduces to $\phi(x,y) = \int_0^z \lambda r_e \delta n_e(x,y,z) dz$, with the electron radius $r_e$ = 2.8179$\times$10$^{-13}$cm. The propagation of a wavefront from a 
source to an observer can be computed  by the Kirchhoff-Fresnel diffraction 
integral (Born \& Wolf 1999). This integral relates the electrical field
 ${\bf E}({\bf x'})$ in the observer's plane to the field  $\bf {E_s}$ 
emitted by a distant point-like source. Classically, 
this integral  yields the diffraction pattern of 
an unperturbed aperture. We modify  this integral by 
adding the Kolmogorov phase field $\phi_K({\bf x})$ of a  
turbulent equivalent screen to model the wavefront corrugations caused by the ionized ISM. The electric field in the observer's plane with the Kirchhoff-Fresnel integral is :  

\begin{equation}
{\bf E}({\bf x'}) = {{\bf E_s}  \over {\lambda |{\bf L}|}} \int_{\mathcal{S}}   
  {{e^{i[{{2 \pi} \over  \lambda }{|{\vec \ell}|} +\phi_K({\bf x})] }} }
dS %  ~~~~~ (1) $$
\end{equation}
 
\noindent where $\lambda$ is the wavelength, 
 ${\bf x'}$ is the observer position vector in a plane parallel
to the sky plane, ${\bf x}$ is the position vector of a 
point in the phase screen plane $\mathcal{S}$ also parallel to the sky plane,  
 ${\bf L}$ is the vector between the observer and the screen planes. 
In addition, $|{\bf x'}|$ and $|{\bf x}|$ are very small with
respect to $|{\bf L}|$ in our application below and this justifies the factor ${ 1  \over {\lambda |{\bf L}|  }}$
is constant in eq. (1).  $|\vec \ell|$ is the full geometric
 path from  the pulsar to the  observer through the screen.  
The geometry is sketched in Fig.~1. 

%\medskip        
%\centerline {[ Place Fig 1 here ]}
%\bigskip

\begin{figure}
\centering
\includegraphics[width=9.2cm]{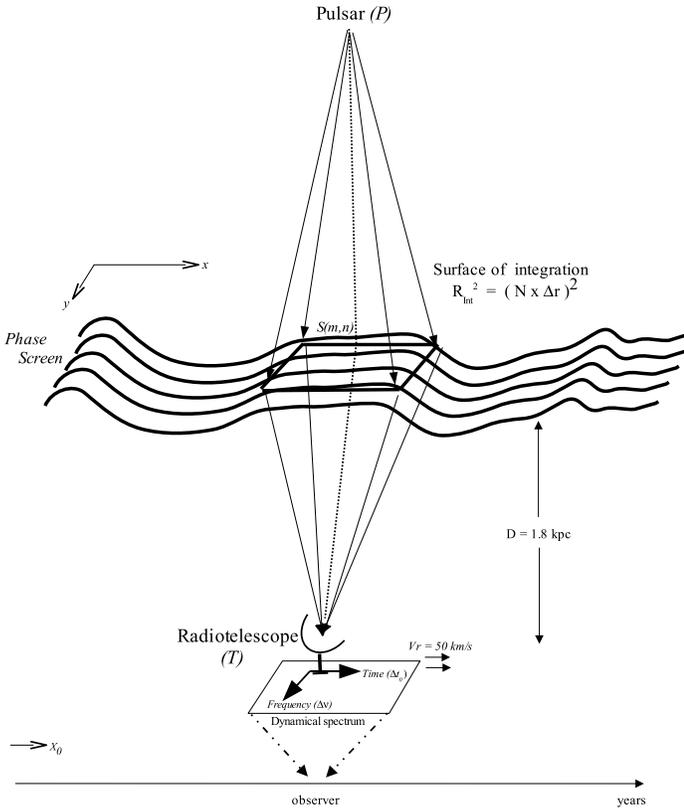}
\caption{Sketch of our computation of the dynamic spectrum of a pulsar  seen through a 2-D turbulent equivalent phase screen. The geometric path $\ell$ in the Kirchhoff-Fresnel integral, eq. (1), is from the point source pulsar $P$  to the radiotelescope position $T$, by crossing the phase screen at the running pixel $S(m,n)$.
}
\label{figure_mafig1}
\end{figure}

The 2-D  power spectrum  of this phase screen 
$ {P_{2\phi}} (q_x, q_y) $  is related to the  
3-D Kolmogorov power spectrum
of the interstellar electronic density 
${\rm {P_{3N}}}(q_x,q_y,q_z=0)$ as shown  in  
Lovelace (1970) and Lovelace et al. (1970) in the thin screen approximation : 
 
\begin{equation}
{P_{2\phi}} (q_x,q_y)  = 2 \pi z (\lambda r_e)^2 
{\rm P_{3N}} (q_x,q_y,q_z=0) %~~~~~ (2) $$
\end{equation}

\noindent $z$ is the screen thickness. The thin screen approximation refers
to a small deviation in the propagation direction of the wave rather
than to a small thickness (Salpeter, 1967). 

In Appendix A, we derive the expression of the phase field $\phi_K (x,y)$ by using 
this relationship and the  definition
of the  2-D average power spectrum  $ P_{2\phi} (q_x,q_y) $  
over the square screen surface $ N^2 {\Delta r}^2 $  recalled here as:

\begin{equation}
P_{2\phi} (q_x,q_y) = { | {\mathcal{F}_{2{\phi}}(q_x,q_y)}|^2  \over { N^2  {\Delta r}^2}} %~~~~~~ (3) $$
\end{equation}

\noindent $\mathcal{F}_{2{\phi}}(q_x,q_y)$
 is the Fourier Transform of the phase field $\phi_K(x,y)$.
In Appendix A, $\phi_K(x,y)$ is cast into a discrete 
form for computation by  FFT.  The observer's plane 
$ {\bf x}= (x,y)$ becomes a grid $(m,n)$ such that $x=m~\Delta r$
  and $y=n~\Delta r$ with the grid step $\Delta r$.
The spatial frequency  plane  $(q_x,q_y)$ becomes the grid $(k,l)$
such that  $q_x=k~dq_x$ and  $q_y=l~dq_y$. The frequency step
$dq_x$  is chosen to be the lowest spatial frequency 
$q_{min,x}  = 1 / N \Delta r$ along $x$ and, similarly,  
$dq_y =  1  / N \Delta r$ along $y$. The grid step $\Delta r$ is 
chosen  to be  $s_0/4$ after the tests 
described in Appendix B and adopting the notation $s_0$ for the coherence length. The maximum spatial frequency $ q_{max} = {1 \over {2~\Delta r}} $ 
 in the integral A.5 satisfies the Nyquist sampling rate. The discrete formula reduced to the case of the square screen is:

\begin{center}
$\phi_K(m,n) = {2 \pi}^{{{(1-\beta)} \over 2}}  ~b~ \lambda r_e \sqrt {z C_N^2}  ~~ {(N \Delta r)}^{-1+\beta/2} $
\end{center}
\begin{equation}
\times ~~\sum_{k=-N/2}^{k=+N/2} ~~ \sum_{l=-N/2}^{l=+N/2}  \left( k ^2 + l^2 \right)^{-\beta/4}  e^{ -{{2 \pi i} \over {N}} ({k m}+ {l n})} ~~ e^{-i\psi(k,l)}
\end{equation}

The grid step $\Delta r$  conveniently appears as part of the
 multiplying factor in eq.~(4) 
so that the double summation of this equation can be computed once 
and  stored in a file 
for use with different  values of  $\Delta r$. This was
particularly useful for the tests below  made with  several phase screens 
constructed by adjusting the multiplying factor.    

We used a complex Hermitian symmetric spectrum 
to make the phase field  $\phi_K(m,n)$ real, {\it i.e.} The  
complex coefficients of eq.~(4) are  conjugate by applying
$\psi(k,l) = - \psi(-k,-l)$  in the
construction of the screen. 
Eq (4) is suitable for computation by FFT 
where the  elements of the input complex array are  the coefficients
$ c(k,l) =  - c(-k,-l) = {\left(k^2 + l^2\right)}^{-\beta/4} e^{\pm i\psi(k,l)} $.  We generate a random phase field  $ \phi_K(m,n) $
by making  the Fourier phase $\psi(k,l)$ a random variable uniformly
distributed over $[0, 2 \pi]$ (Rice 1944 p. 287).  
We had to make the adjustment factor $b = 2 $ in eq.(4)
so that the phase structure function $D_{\phi}(r)$ 
computed directly from the screen yields a coherence length
that matches the theoretical value $s_0$. Coles et al. (1995) have devised a method to randomize a phase screen. Instead of our complex numbers $b e^{-i\psi(k,l)}$ in eq. (4), with the random phase $\psi(k,l)$
described above, Coles et al. (1995) set complex random numbers 
made of independent zero-mean Gaussian random variables $x$ and $y$ with variance $\sigma^2 $ for their real and imaginary parts.
In polar coordinates, these  variables $x$ and $y$ have magnitudes following a Rayleigh probability distribution
and phases distributed uniformly over $[0, 2 \pi]$ as for our $\psi(k,l)$ (see Thompson et al, 1986, p.259-260 and reference to Papoulis, 1956).
 We have verified in generating independent zero-mean Gaussian random variables $x$ and $y$
 of various  $\sigma$ that  the mean amplitude of the corresponding 
 Rayleigh probability distribution for $\sigma=1.6$ is 2, {\it i.e.} the value of our b factor. Thus, 
the method we use  to generate the random complex numbers in eq. (4) is a satisfactory approximation of the formal 
synthesis of the random phase field $\phi_K$ in Coles et al. (1995).

%The two solutions for the choice of random complex numbers in eq. (4) can be made consistent this way.  

The intensity of the pulsar  $i({\bf x', \lambda})= {\bf|E|}^2({\bf x', \lambda})$ 
at  wavelength $\lambda$  is computed with the discrete 
form of the  Kirchhoff-Fresnel integral where
the gridded phase screen $\phi_K(m,n)$ substitutes 
the continuous Kolmogorov phase $\phi_K({\bf x})$. 
The electric field (eq. 1) at the grid position $p$ of the radiotelescope  
and  wavelength $\lambda$  is  :

\begin{equation}
{E}(p,\lambda)=   {{E_s {dx_s}^2} \over {\lambda L }} \sum_{n=1}^{n=H} ~~
\sum_{m=1}^{m=H} e^{i[{{2 \pi} \over {\lambda}} {\ell}(p,m,n) 
+ \phi_K(m,n)]}
\end{equation}

\noindent where the geometric path $ \ell(p,m,n)$ is from the point source pulsar $P$ through the phase screen at the running pixel $S(m,n)$ to the 
radiotelescope position $T$ along the x-axis $(p)$ in Fig.~1.
We have not approximated  this path by a power series in our code so that we assumed neither
 the Fraunhofer approximation nor the Fresnel approximation but carried out the full computation
of $ \ell(p,m,n)$. 
The parameter $dx_s$  is the spatial resolution of the phase screen
 which does not have to be equal 
to the grid step $\Delta r$  controlling  
the limits $q_{min}$  and  $q_{max}$ of the Kolmogorov spectrum
$P_{3N}$.  Since eq~(5) is the discrete form of  eq~(1), 
the parameter $H$ of eq~(5) controls the size of the integration surface 
noted $\mathcal{S}$ in eq~(1); $\mathcal{S}= (H \times dx_s)^2$.

\section{Construction of a Large Square Kolmogorov Phase Screen}\label{kolm} 

We have computed a relatively large Kolmogorov phase screen for the pulsar PSR1937+21
 with eq.~(4) on an alpha server assuming  turbulence 
strength $C_n^2 = 10^{-3} {\rm m}^{-20/3}$ and  screen thickness $z$ 
to be the pulsar distance (3.6 kpc) (this assumption is discussed below). The size of the screen is
$N=2^{17}$, {\it i.e.} there are $\sim$ 131k x 131k pixels. Obviously, this computation  could not be done 
in a single 2-D FFT and we had to resort to multiple 1-D FFT's. 
 Each line of the input complex array, coefficients $c(k,l)$ of eq.(4),  was first transformed 
by a $2^{17}$ point 1-D FFT along the $x$-axis.
Then, each resulting column was transformed by a $2^{17}$  point 1-D 
FFT along the $y$-axis. 
The construction of this large square phase screen 
 took 10 days on our workstation and needs 68~GigaBytes of disk space to
store all the phases but the peak storage during the computation was 200 GigaBytes. 

With the assumed  turbulence 
strength $C_n^2 = 10^{-3} {\rm m}^{-20/3}$ and thickness
$z = 3.6$ kpc, the  
coherence length in the screen is $s_0=2.66~10^7$ m at 1.41 GHz as calculated 
from the   phase structure function $D_\phi(s_0)=1~{\rm Rd}^2$ 
with the theoretical expression  
$D_\phi(s) = 8 \pi r_e^2 \lambda^2 C_N^2 z f(\alpha) (\alpha+1)^{-1} s^{\alpha}$, where numerically $f(\alpha)$ = 1.12 for $\alpha$=5/3 (Armstrong, Rickett \& Spangler 1995). 
To qualify our phase screen, we have computed  the phase  structure functions
 $D_\phi(s)$ along multiple  $x$-lines through the screen and, 
orthogonally, along $y$. 
From the log-log plots of these functions 
(examples in Fig.~2), we made  linear least-square-fits and found slopes of 
$1.68 \pm 0.10$ for $x$ and $1.64 \pm 0.1$ for $y$. The theoretical
slope of $D_\phi(s)$  is $\alpha = 5/3$
for an isotropic turbulent medium with the exponent $\beta = 11/3$
in $P_{3N}$.  
We have also measured the coherence lengths  from these
 phase structure functions:  $r_{coh} = 2.4 \pm 0.6 ~10^7$ m  
along $x$ and  $r_{coh} = 3.2 \pm 0.6 ~10^7$ m along $y$. 
These values are consistent with the value $s_0 = 2.66~10^7$ m predicted by the  
structure function $D_{\phi}(s_0)=1$ Rd. This is also consistent with the value obtained 
from the equation $r_{coh} = V/t_d \simeq$2.1 10$^7$ m, where $V = 50$ km/s and $t_d = 7$ min 
is the observed diffractive timescale toward the pulsar B1937+21 (Ryba 1991). We note that 
it is coincidental that  $s_0$ has the expected value by using the conventional value of $C_n^2$ for PSR1937+21
and our assumption above
that the  thickness $z$ equals the pulsar distance.

%\medskip  
%\centerline {[ Place Fig 2 here ]}
%\bigskip

\begin{figure}[t]
\centering
\includegraphics[width=6.5cm,angle=-90]{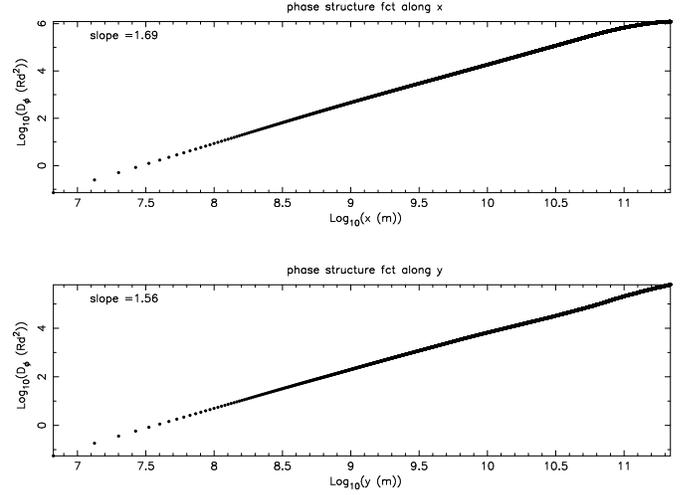}
\caption{Examples of the phase structure functions $D_{\phi}(x)$ and  $D_{\phi}(y)$ computed for 1-D cuts along the x and y axis across the whole  131k $\times$ 131k phase screen $\phi_K(m,n)$ constructed with $C_n^2 = 10^{-3}$, $z = 3.6$ kpc  and $\Delta r = s_0$. The slopes of  these structure functions  measured from the log-log plots are satisfactorily close to the theoretical value $\alpha = 5/3$ for $\beta = 11/3$ of the Kolmogorov spectrum $P_{3N}$.}
\label{figure_mafig2}
\end{figure}

\section{Results: Simulation of the Scintillation of the \\
Pulsar B1937+21 at 1.41 and 1.7~GHz}\label{res} 

Prior to the full scale simulations, we studied
the convergence of the computation of the dynamic
spectra that depends on four  parameters : the size $S$ of the 
integration surface of eq.~(1), the spatial 
resolution $dx_s$ to read the phase screen file, 
the grid step $\Delta r$  in the  phase screen  and the size $N$ of the phase screen
 (eq.~4) which control the 
minimum and maximum spatial frequencies $q_{min}= {1 \over {N \Delta r}}$
and $q_{max}= {1 \over {2\Delta r}}$ of the Kolmogorov spectrum 
$P_{3N}$. The parameter $N$ was fixed to $2^{17}$, limited only by the maximum disc space
available with our computer to store the phase screen,  and this corresponds
to a period of 6 months if the phase screen drifts across the observer line of sight at $V = 50 $ km/s.  
We have empirically determined the other three parameters by a sequence 
of tests that are described in detail in Appendix B.

   In Fig.~3, we present  the intensity of  the pulsar B1937+21 
   simulated during this  period of  six months at 1.41 and 1.7~GHz.
   The intervening phase screen  $\phi_K(m,n)$  is 
   constructed with $C_n^2$ = 10$^{-3}$m$^{-20/3}$ and  $z$ = 3.6 kpc that are  the  parameters of 
   B1937+21 (Ryba 1991; Johnston, Nicastro \& Koribalski 1998), 
   $N= 2^{17}$ and $\Delta r = s_0/4$
   in eq.~(4) based on our conclusion in the second test (Appendix B).
   The coherence length in the phase screen is $2.6~10^{7}$~m thus the lower and upper limit scales of the Kolmogorov
   spectrum $P_{3N}$ are  $q_{min}= {1 \over {2^{17} \times 2.6~10^{7}}} = {1 \over {3.4~10^{12}}}~m^{-1} $
   and $q_{max}= {1 \over {2 \times 2.6~10^{7}}}~m^{-1}$.
   We adopt the integration surface $S=(4 r_S)^2$ to compute
   eq.~(5) from the conclusion in the first test (Table 1). We use
   the resolution $dx_s = s_0/4$ (Table 2) to read the phase 
   screen file into eq.~(5).  This resolution is 
   conservative relative to our conclusion in the third test;
   it was dictated for algorithmic peculiarities to the expense 
   of computing speed. A dynamic spectrum
   was computed in these conditions every other 1.25 days. 
   Averaging these dynamic spectra provides the intensity measured by the telescope
   over the integration time 70 min and the bandwidth 8.8 MHz,
   sampled over $32 \times 32$ time bins and frequency channels 
   in our computation. The velocity of the screen
   is $V =50$ km/s. This simulation took twice 1.5 months 
   on our alpha server. 

%\medskip  
%\centerline{[ Place Fig 3 here ]}
%\bigskip

\begin{figure*}
\centering
\includegraphics[width=12cm,angle=-90]{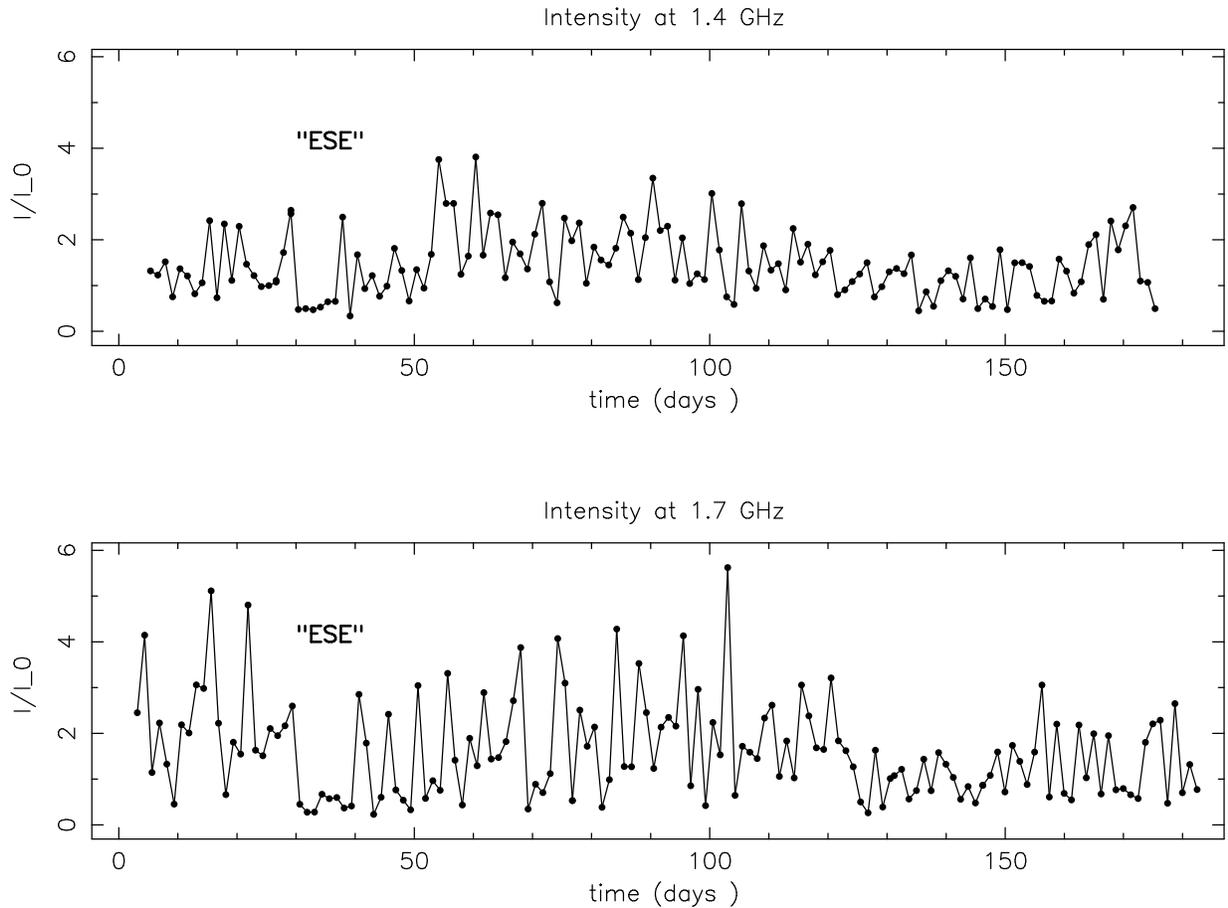}
\caption{Intensity of B1937+21 simulated  at 1.41 GHz and 1.7 GHz. 
The intervening phase screen is constructed  with
 $C_n^2 = 10^{-3}$, $z = 3.6$ kpc, $N=2^{17}$ and $\Delta r = s_0/4$
in eq.~(4) and $S = (4 r_S)^2$ and $dx_s = s_0/4$ in  eq.~(5).  Between days 30 and 40, there is an
 event occurring simultaneously at 1.41 and 1.7 GHz which 
is alike the behavior of the flux density recorded in direction of B1937+21
at Nan\c cay in October 1989 and interpreted as an Extreme Scattering Event. 
The mean intensities $<I>$ are 1.46 and 1.60 and the modulation indexes
$m$ are 0.49 and 0.67 for these two series.}
\label{figure_mafig3}
\end{figure*}

   Overall, Fig.~3 shows the typical behavior of flux density recorded  
   for pulsars; for instance B1937+21 at Nan\c cay at 1.41 and 1.7 GHz
   in   Figs.~1 and 2 
   of LRC98. In addition, 
   there is the interesting  feature labeled ``ESE'' and present simultaneously 
   at 1.41 and 1.7~GHz in Fig.~3 from day 30 to day 40. The significant
   drop of flux density (60 \% below the mean value) and the low rms during this ESE period (rms=0.10 during the ESE versus rms=0.71 off ESE) for these 
   10 days is alike
   the ESE observed toward B1937+21 in October 1989 (Cognard
   et al. 1993). In this previous paper, we suggested
   this phenomenon was caused by a discrete cloud of plasma
   following the standard interpretation of Fiedler at al. (1987).  
   Our simulation shows instead that this event can arise 
   naturally because of the turbulence in the ionized interstellar
   medium as conjectured. We have labeled this event ``ESE'' in Fig.~3. The ratio of the duration
   of this event to the length of our simulation is $\sim \rm 10 days / 180 days$ $\simeq$ 5\%.
 Although this might be fortuitous, this percentage is consistent with the observations of B1937+21 at Nan\c cay  for which the rate of occurrence of ``ESE's''
is 4\% (LRC98). Figure 4 shows the cross correlation function of the flux densities at 1.41 and 1.7 GHz (Figure 3). The zero-lag value $\simeq$0.8 is consistent with the observed value 0.93$\pm$0.05 (LRC98). 

%\medskip  
%\centerline{[ Place Fig 4 here ]}
%\bigskip

\begin{figure}
\centering
\includegraphics[width=6cm,angle=-90]{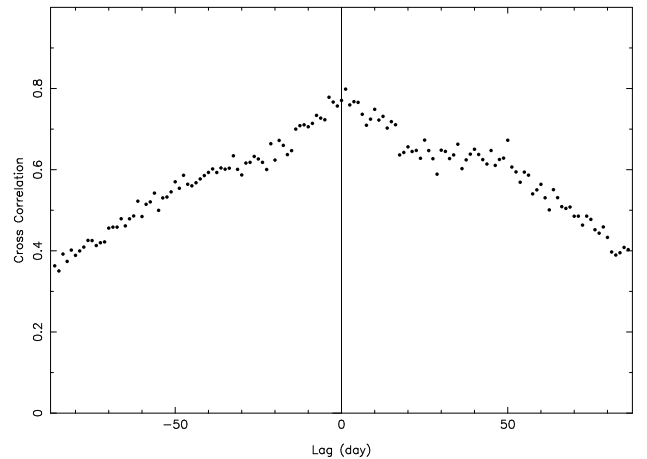}
\caption{Cross-correlation function of the flux densities at 1.41 GHz and 1.7 GHz of Figure 3.}
\label{figure_mafig4}
\end{figure}

The duration of the event labeled ``ESE'' in our simulation is
comparable to the event observed at Nan\c cay in October 1989 (12 days)
but is short compared to three others of duration 1-3 months identified 
in the radio light curves of B1937+21 simultaneously at 1.41 and 1.7~GHz 
 by LRC98. The extraordinary focusing of pulsar waves is caused by the large scale fluctuations of the electronic density in the interstellar medium.
The largest fluctuation scale in our simulation is 
$2^{17} \times 2.6~10^{7}/4 \sim  10^{12}$~m while  
the ionized interstellar medium is made of fluctuations
up to $10^{18}$~m (Armstrong, Rickett and Spangler 1995).
We expect that longer ``ESE''s require to  include fluctuations of larger scales  in our simulation. We plan to extend this calculation on a supercomputer. We note here  that the relatively long events labeled ``ESE'' in Hamidouche, Lestrade, and Cognard (2002) are from a simulation that
was improperly done with an integration surface as small as $S = (0.7 r_S)^2$ chosen before we made the tests of the present paper.     

Finally, we note that the anti-correlation coefficient -0.62 between  flux density and TOA in the simulation 
conducted with a rectangular phase screen is reported in Hamidouche (2003), and is comparable to the correlation coefficient  -0.43 to 
-0.73 depending on data segments of the Nan\c cay observations in LRC98. We have not calculated this correlation in the
final simulation.

\section{Conclusion}\label{conc} 

 We have presented the simulation of the scintillation 
of pulsars carried out within the frame of Physical Optics by extending 
the seminal work by Cordes, Pidwervetsky \& Lovelace (1986).
We have built a large square Kolmogorov phase screen of 
$131 {\rm k} \times 131{\rm k}$
pixels on an alpha server and shown how to set properly 
the parameters of the Kirchhoff-Fresnel integral by conducting several tests. 
Our final simulation  was done for the condition of turbulence known
in the direction of the pulsar B1937+21 
($C_n^2 \sim  10^{-3}$ and distance  $z = 3.6$ kpc) and the radio light curves
were  generated  at 1.41 and 1.7 GHz for a period of 6 months. 

These two light curves exhibit 
simultaneous variations  at 1.41 and 1.7~GHz
that are alike the ``ESE'' observed at these frequencies 
at Nan\c cay in October 1989
and that lasted $\sim 10$ days. Our simulation shows
that this observed event can be caused naturally by the turbulence in the
ionized interstellar medium instead of invoking 
the crossing of a discrete over pressured ionized cloud
on the line of sight as in the model of Fiedler et al. (1987). 
We think that longer events could occur by including the electronic density fluctuations 
of larger scales in the construction of the Kolmogorov phase screen used in the simulation. 

%% On the other hand, another way to simulate the phase screen is using another concept to
%%  generate the phase of the simulated Kolmogorov phase screen. Instead of adding a a unity
%%  amplitude and a uniformly randomly distributed phase, one would use a non-unity amplitude
%%  phase where its real and imaginary parts are zero mean Gaussian random variables with 
%% a unity variance (Coles et al. 1995). In our next simulation, we will try both methods 
%% to compare the results of the scintillation and the rate of production of the 
%% ESE (Hamidouche \& Lestrade 2007). We will conduct the simulation over 10 years, 
%% including larger fluctuation structures in our screen, and compare directly to the 10 years observations of the pulsar B1937+21 (LRC98).

Finally, we note that Deshpande (2000) stresses that the opacity differences in HI and other species  measured over a transverse separation $x_0$
result from all scale  of the 3-D power spectrum of the
opacity fluctuations while they are currently, and erroneously 
in his opinion,  interpreted as  over pressured and overdensed
cloudlets of size $x_0$. Our simulation of ``ESE's'' caused by the turbulent
ionized interstellar medium strengthens this opinion for the neutral gas.
\bigskip

\indent {\bf Acknowledgments:} We are grateful to our referee Prof. Barney Rickett who  carefully read  the manuscript and provided
many valuable suggestions to improve the paper.  
We are in debt to the legacy of Max Born and Emil Wolf. We would like to thank Fran\c coise Combes and
 Fran\c cois Lefeuvre for useful discussions. M.H. would like to thank, Richard Crutcher, the director of the Laboratory for Astronomical Imaging at the University of Illinois. 
\bigskip

\noindent {\bf References:}

\bibl
Armstrong, J.W., Rickett, B.J. \& Spangler, S.R., 1995, ApJ,
 443, 209
\filbreak

\bibl
Blandford, R., Narayan, R., 1985, MNRAS, 213, 591.
\filbreak

\bibl
Born, M, Wolf, E., 1999, {\it Principles of Optics}, 7th edition, (Cambridge
University Press)
\filbreak

\bibl
Cognard, I., Bourgois, G., Lestrade, J-F, Biraud, F., Aubry, D., Darchy, B., 
Drouhin, J-P, 1993, Nature, 366, 320
\filbreak

\bibl
Coles, W.A., Filice, J.P., Frehlich, R.G., Yadlowsky, M., 1995, Applied Optics, 34, 2089
\filbreak

\bibl
Cordes, J.M., Weisberg, J.M, Boriakoff, V., 1985, ApJ, 288, 221-247.
\filbreak

\bibl
Cordes, J.M., Pidwervetsky, A., Lovelace, R.V.E. (1986), ApJ, 310, 737.
\filbreak

\bibl
Deshpande, A., 2000, MNRAS, 317, 199-204.
\filbreak

\bibl
Fiedler, R.L.,  Dennison, B., Johnston, K.J, Hewish, A., 1987, 
Nature, 326, 675
\filbreak

\bibl
Fiedler, R.L., Dennison, B., Johnston, K.J., Waltman, E.B., Simon, R.S., 
1994, ApJ, 430, 581.
\filbreak

\bibl
Goodman, J.W., 1968, {\it Introduction to Fourier Optics} (New York: McGraw-Hill)
\filbreak

\bibl
Gupta, Rickett, B.J., Coles, W.A., 1993, ApJ, 403, 183-201
\filbreak

\bibl
Gupta, Rickett, B.J., Lyne, A., 1994, MNRAS, 269, 1035
\filbreak

\bibl
Gwinn, C.R., Britton, M.C., Reynolds, J.E., Jauncey, D.L., King, E.A., 
McCulloch, P.M., Lovell, J.E., Preston, R.A., 1998, ApJ, 505, 928-940.
\filbreak

\bibl
Hamidouche, M., Lestrade, J-F, Cognard, I., 2002, in Proceedings `` 
Scientific Highlights 2002 of the Soci\'et\'e Fran\c caise d'Astronomie 
et d'Astrophysique'', Paris, June 2002,
ed. F. Combes,  D. Barret, F., EDP Sciences Conference Series 
in Astronomy, 29. 
\filbreak

\bibl
Hamidouche, M., 2003, ``Simulation of the Interstellar Scintillation of radio Pulsars. Characterization of Extreme Scattering Events toward the pulsar B1937+21.'', PhD Thesis, University of Orl\'eans
\filbreak

\bibl
Hill, A.S., Stinebring, D.R., Asplund, C.T., Berwick, D.E., Everett, W.B., Hinkel, N.R., 2005, ApJ, 619, L171-L174
\filbreak

\bibl
Johnston, S., Nicastro, L. \& Koribalski, B., 1998, MNRAS, 297, 108-116
\filbreak

\bibl
Lazio, T.J.W., Waltman, E.B., Ghigo, F.D., Fiedler, R.L., Foster, R.S., Johnston, K.J., 2001, ApJS, 136, 265 
\filbreak

\bibl
Lestrade, J.-F., Rickett, B.J., Cognard, I. 1998, A \& A, 334,
1068
\filbreak

\bibl
Lovelace, R.V.E., 1970, PhD Thesis, Cornell University
\filbreak

\bibl
Lovelace, R.V.E., Salpeter, E.E., Sharp, L.E., Harris, D.E., 1970, ApJ, 159, 1047-1055
\filbreak

\bibl
Maitia, V., Lestrade, J-F, Cognard, I., 2003, ApJ, 582, 2, 972-977.
\filbreak

\bibl
Narayan, R., 1992, Phil. Trans. R. Soc. Lond. A, 341, 151
\filbreak

\bibl
Papoulis, A., 1965, Probability, Random Variable and Stochastic Process, McGraw-Hill
\filbreak

\bibl
Rice, S.O., 1944, {\it Mathematical Analysis of Random Noise}, Bell System 
Tech. J., 23, 282
\filbreak

\bibl
Rickett, B.J., 1977, ARA\&A,  15, 479
\filbreak

\bibl
Rickett, B.J., 1988, in ``Radio wave scattering in the interstellar medium'', 
 Proceedings of the AIP Conference, San Diego, CA, Jan. 18-19, 1988,
 New York, American Institute of Physics, 1988, p. 2-16. 
\filbreak 

\bibl
Rickett, B.J., 1990, ARA\&A,  28, 561
\filbreak

\bibl
Romani, R.W., Blandford, R.D., Cordes, J.M., 1987, Nature, 328, 324
\filbreak

\bibl
Ryba, M.F., 1991, ``High Precision Timing  of Millisecond Pulsars'', PhD Thesis Princeton University
\filbreak

\bibl
Salpeter, E.E., 1967, ApJ, 147, 433
\filbreak

\bibl
Stinebring, D.R., Smirnova, T.V., Hankins, H., Hovis, J.S., Kaspi, V.M.,   
Kempner, J.C., Myers, E., Nice, D.J., ApJ, 539, 300-316.
\filbreak

\bibl
Thompson, A.R., Moran, J.M., Swenson, G.W., 1986, Interferometry and Synthesis in Radioastronomy, 1rst edition, Wiley-Interscience
\filbreak

\clearpage

%\onecolumn

{\bf \centerline{Appendix}}
\appendix
\section{Construction of a Kolmogorov Phase Screen}\label{apA}

The definition of the 2-D phase power spectrum of the phase field
$\phi(x,y)$ averaged over the rectangular screen surface $N M \Delta r^2$ is :

\begin{equation}
P_{2\phi} (q_x,q_y) = { | { {\bf \mathcal{F}}_{2{\phi}}(q_x,q_y)}|^2  \over { N M {\Delta r}^2}}
\end{equation}

where $\mathcal{F}_{2{\phi}}(q_x,q_y)$ is the Fourier Transform of $\phi(x,y)$.
Hence, the module of the  Fourier component for the frequencies $(q_x,q_y)$ is:
 
\begin{equation}
|\mathcal{F}_{2 \phi}(q_x,q_y)| = \sqrt { N M \Delta r^2 P_{2\phi}(q_x,q_y)}
\end{equation} 

\noindent Lovelace (1970) and Lovelace et al. (1970) have established:

\begin{equation}
{P_{2\phi}} (q_x,q_y)  = 2 \pi z (\lambda r_e)^2 
{\rm P_{3N}} (q_x,q_y,q_z=0)
\end{equation}

\noindent  Hence, the complex Fourier components of the phase field
$\phi(x,y)$ are: 

\begin{equation}
\mathcal{F}_{2\phi}(q_x,q_y) = \sqrt{2 \pi}  \lambda r_e \sqrt { N M \Delta r^2 z C_n^2 
 (q_x^2 + q_y^2)^{-\beta/2}}   ~~ e^{-i\psi(q_x,q_y)}
\end{equation}

that must satisfy the complex Hermitian symmetric spectrum with
the Fourier phase $\psi(q_x,q_y) = - \psi(-q_x,-q_y)$ 
to make the phase field $\phi(x,y)$  real. $\psi(q_x,q_y)$ is a random 
variable uniformly
distributed over $[0, 2 \pi]$ as prescribed by Rice (1944, p.287). 
$C_n^2$ is the turbulence strength and $z$ is the propagation length.  

The phase field $\phi(x,y)$ can be computed by the inverse Fourier Transform 
of $\mathcal{F}_{2 \phi}(q_x,q_y)$. Although $P_{3N}$ is a power-law,
the Fourier integral is finite because there is an  outer scale in the 
ionized interstellar medium  that makes  
$P_{3N}$ becomes zero for small $q$ rather 
than infinity when  $q \rightarrow 0 $. Also, we point out that 
there is the factor  $(2 \pi)^{-2}$ in  this integral to be
consistent with  the Fourier Transform definition used to define $P_{3N}$ 
in Rickett (1977, eq. 6) and used to demonstrate  
the Lovelace relationship (Lovelace 1970, eq. 36). This inverse Fourier
Transform is:
 
\begin{center}
$\phi(x,y) = (2 \pi)^{-2+{1 \over 2}}  \lambda r_e \Delta r 
\sqrt { N M  z C_N^2}~~ \times$
\end{center}
\begin{center}
$\Big[ \int_{-q_{x,max}}^{-q_{x,min}} 
\int_{-q_{y,max}}^{-q_{y,min}} (q_x^2 + q_y^2)^{-\beta/4} 
e^{-i\psi(q_x,q_y)} e^{-i(q_x x + q_y y)}  dq_x dq_y ~~ + $
\end{center}
\begin{equation}
\int_{+q_{x,min}}^{+q_{x,max}} \int_{+q_{y,min}}^{+q_{y,max}} (q_x^2 + q_y^2)^{-\beta/4} 
e^{-i\psi(q_x,q_y)} e^{-i(q_x x + q_y y)}  dq_x dq_y  \Big] 
\end{equation}

This integral was split into two parts to avoid the singularity 
of the power-law function.  In eq.~(A.5) for the rectangular 
 screen case, the minimum frequencies are 
$q_{x,min} = {1 \over {N \Delta r}}$ and 
$q_{y,min} = {1 \over {M \Delta r}}$.

% The discrete form for the phase field requires the substitutions  
%  $ x=m  ~ \Delta r $, $y = n ~ \Delta r$,  
% $q_x= k ~ dq_x = 2 \pi k/ N \Delta r$ 
% and $q_y= l ~ dq_y = 2 \pi l/ M \Delta r$ with  $dq_x = q_{x,min}$ 
%  and   $dq_y = q_{y,min}$ : 

% \begin{equation}
% \phi(m,n) =  {2 \pi}^{{{(1-\beta)} \over 2}} \lambda r_e \sqrt { {z C_N^2} \over { N M}} ~~ 
% {\Delta r}^{-1+\beta/2} ~~ \times ~~~~~~~~~~~~~~~~~~~~~~~~~~~~ ~~~~~~~~~~~~~~~~~~~~~~~$$
%       $$ ~~~~~~~~~ \sum_{k=-N/2}^{k=+N/2} ~~~ \sum_{l=-M/2}^{l=+M/2} 
% \left( \left( k \over N \right)^2 + \left( l \over M \right)^2 \right)^{-\beta/4}  e^{-2 \pi i ({{k m} \over N}+ {{l n } \over M}) } ~~ e^{-i\psi(k,l)}
% \end{equation}

% This form is suitable for a FFT subroutine. This rectangular phase
% screen can be turned into a square screen for our application by
% setting $N$ = $M$. 

\clearpage

\section{Tests for the Computation of the Dynamic Spectra}\label{apB}

Prior to the full scale simulations, we studied
the convergence of the computation of the dynamic
spectra that depends on four  parameters : the size $S$ of the 
integration surface $\mathcal{S}$ of eq.~(1), the spatial 
resolution $dx_s$ to read the phase screen file, 
the grid step $\Delta r$ in the screen phase, and the size $N$ of the phase screen
 (eq.~4). These latter two parameters control the 
minimum and maximum spatial frequencies $q_{min}= {1 \over {N \Delta r}}$
and $q_{max}= {1 \over {2\Delta r}}$ of the Kolmogorov spectrum 
$P_{3N}$. The parameter $N$ is fixed to $2^{17}$ in our computation and 
we have empirically determined  the other three parameters.

As a first test, we study the effect of $S$ of eq.~(1)
while keeping the parameters $\Delta r$ and $dx_s$ fixed and equal to $s_0$.
We express the dimension of the integration surface $S$ in terms of the scattering radius $r_S$ since, as shown below, the required $S$ to make the dynamic spectra numerically convergent is a few times this scale. In fact, Coles et al. (1995) have demonstrated that the integration surface in the phase screen has to be larger than the scattering disk. The physical idea is that the scattering disk is the approximate area in the phase screen irradiating a single location in the observing plane.     
We  chose the following  sizes $S$ for our test: $S = (1 r_S)^2$, $S = (2 r_S)^2$, $S = (4 r_S)^2$, $S = (6 r_S)^2$,
 $S = (8 r_S)^2$, $S = (10 r_S)^2$.
For each size $S$, we compute the whole series of dynamic 
spectra at 1.41 GHz by shifting $\mathcal{S}$ by 2.5 days across 
the screen using V=50 km/s. Each dynamic spectrum
is  70 min x 8.8 MHz in size, similarly to observations 
of B1937+21 at Nan\c cay, and sampled over 8 x 8 pixels. 
For each  series, we  derive the mean intensity $<I>$ and
the modulation index $m = {I_{\rm rms} \over < I >}$ in Table~B.1. This  
table shows that  these two indicators  
converge  when $S$ reaches $(4 r_S)^2$.    
In Fig.~B.1,  we show only three  series for clarity; $S= (2 r_S)^2$, $S= (4 r_S)^2$ and $S= (10 r_S)^2$ to depict the convergence process. 
In addition, we illustrate this convergence process in Fig.~B.2 by 
showing dynamic spectra computed at the same position in the observer 
plane but for the 6  integration surfaces mentioned above.
From this first test summarized by Table~B.1, Figs.~B.1 and B.2, 
we conclude that  convergence of  dynamic spectra starts  for  $S= (4 r_S)^2$ since  improvement in the average intensity and index $m$  are less than 10 \% for larger surfaces $S$.

\begin{center}

\begin{table} [h] 
 \caption[]{Sensitivity of the averaged intensity  and modulation index to the integration surface at 1.4GHz $^{\mathrm{a}}$}

 $$ 
         \begin{array}{p{0.5\linewidth}lccc}
\hline\hline
            \noalign{\smallskip}
%%%\begin{center}
%%%\begin{tabular}{|l|c|c|c|} \hline
\rm {Integration}    & \rm { Nber~of} &  <I>  & \rm { Index}~$m$  \\
\rm {surface}~$S$    & \rm { intensities}    &         &      \\
\noalign{\smallskip}
            \hline
            \noalign{\smallskip}
$(1 r_S)^2$    &  303     &  0.28   &    0.95    \\
$(2 r_S)^2$    &  302     &  0.83   &    0.80    \\    
$(4 r_S)^2$    &  299     &  1.39   &    0.63    \\
$(6 r_S)^2$    &  296     &  1.52   &    0.59    \\
$(8 r_S)^2$    &  292     &  1.55   &    0.58    \\
$(10 r_S)^2$   &  289     &  1.56   &    0.58    \\
 \noalign{\smallskip}
            \hline\hline
         \end{array}
     $$ 
\begin{list}{}{}
\item[$^{\mathrm{a}}$] $\Delta r = dx_s = s_0$ kept fixed.
\end{list}
   \end{table}

%%%%%\end{tabular}
%%%%%\end{center}
%%%%%\caption{Averaged intensity  $<I>$ and modulation index $m$
%%%%%for each  series computed with the fixed increments $\Delta r = dx_s = s_0$
%%%%%but with  the increasing integration surface size $S$.   
%%%%% The second column of this table reflects
%%%%%the numbers of intensities computed over  2 years  $({{N \times \Delta r} \over V})$. They are 
%%%%%slightly different because the largest the surface $S$ is,  
%%%%%the fastest it reaches the border of the screen.}
%%%%%\end{table}

\end{center}

As a second test, we study the impact of the grid step $\Delta r$ in the screen phase
while keeping fixed  $S= (4 r_S)^2$ and $dx_s = s_0$. 
 The parameter $\Delta r$ is 
directly related  to the minimum and maximum spatial 
frequencies $q_{min}= {1 \over {N\Delta r}}$ and 
$q_{max}= {1 \over {2\Delta r}}$ 
of the Kolmogorov spectrum $P_{3N}$ as already mentioned. Note that
 $\Delta r$  appears conveniently
in the multiplying factor of the phase screen 
$\phi_K(m,n)$ in eq.~(4) and hence it can  be easily changed to any value 
to modify $q_{max}$ and, correlatively, $q_{min}$.  
The parameter $N$ is not amendable after the double summation of eq.~(4)
has been calculated and stored in a computer file. In principle, 
we would have liked to tune  $N$ in order 
to keep  $q_{min}= {1 \over {N \Delta r}}$ unchanged while $\Delta r$ is adjusted to increase
 $q_{max} = {1 \over {2 \Delta r}}$. One expects that the high frequency part 
of the Kolmogorov spectrum becomes insignificant in shaping  the
dynamic spectrum when $q_{max}$
is sufficiently high while the low frequency part
modifies it drastically.  This test is difficult to implement in
practice because it requires computing several screens 
with different $N$  while keeping the same random phases $\psi(k,l)$ for the
lower part of the spectrum. Instead of this approach, we simulate
the effect  by superimposing  a corrugated surface of period   $q^{-1}$ 
and  amplitude $\delta \phi$ 
 upon  the original screen. The aim is to seek   
 which  perturbating surface ($q, \delta \phi$)
 degrades significantly the  dynamic spectrum when compared
to the one  computed with  the original phase screen.
This test  covers the following corrugated surfaces :
 $ \delta \phi = 101^{\circ}$ and  $q= {1 \over {2 \times 2 s_0}}$ ;
 $\delta \phi  = 57^{\circ}$and  $q={1 \over {2 s_0}}$;
 $\delta \phi=32^{\circ}$ and  $q={1  \over {2 s_0/2}}$;
 $\delta \phi = 18^{\circ}$ and  $q={1  \over {2 s_0/4}}$;
 $\delta \phi = 10^{\circ}$ for  $q={1  \over {2 s_0/8}}$. 
These amplitudes $\delta \phi$   are derived from 
the phase structure function $D_{\phi}(s)$
for the separations  $s=2~s_0$, $s=s_0$, $s=s_0/2$, $s= s_0/4$ and 
$s= s_0/8$.
Fig.~B.3 shows the comparison of the dynamic spectra computed
with these five perturbations ($ q,  \delta \phi$)  
as well as the dynamic spectrum simulated with the original screen ($\delta \phi = 0$) constructed  with
 $C_n^2 = 10^{-3}$, $z = 3.6$ kpc, $N=2^{17}$  and $\Delta r = s_0/4$.  
The morphology and averaged intensities of these dynamic spectra indicate that 
perturbations are significant above ($\delta \phi = 32^{\circ}, q={1  \over {2 s_0/2}}$),
{\sl i.e.} $q_{max}$ must be at least as high as 
${1  \over {2 s_0/2}}$, {\it i.e.} $\Delta r \le s_0/2$,  for convergence
of the computation. 

%%\medskip  
%%\centerline {[ Place Fig B3 here ]}
%%\bigskip

We complement this test done at a single position of the observer 
by simulating  three intensity series  
across the full screen with the grid step $\Delta r = dx_s$  set to $1 s_0$,  $s_0/2$ and $s_0/4$.
The dynamic spectra are sampled over $8 \times  8, 16 \times 16, 
32 \times 32 $
pixels, respectively, to synthesize  the same 70 min x 8.8 MHz domain.
 Table~B.2 shows that the 
mean intensity $<I>$ and modulation index $m$  are within 10 \% over these three cases.
From this second test   
we conclude, somewhat conservatively, that  $\Delta r = s_0/4$ is 
necessary for the convergence of the computation, {\it i.e.} the Kolmogorov spectrum $P_{3N}$ must include $q_{max} = {1 \over {2~s_0/4}}$.

%%%%\begin{center}
\begin{table} [h]
\caption[]{Sensitivity of the averaged intensity and modulation index to $q_{max}$   $^{\mathrm{a}}$ }
\vspace{0.2cm} 
%%%%\begin{center}
\begin{tabular}{lccccc} \hline\hline
   $\Delta r$   &   $q_{max} = {1 \over { 2~\Delta r}}$    & { Sampling}        & { Nb~of }       &  $<I>$       &  {Index~$m$}       \\
                &                                         &  {of~dyn.~sp.}     & {intensities}     &            &                   \\
\hline
  $s_0$           &   $  {1 \over {2~s_0  } } $    &   $  8 \times  8  $    &    299        &  1.39      &    0.63       \\
  $s_0/2$         &   $  {1 \over {2~s_0/2} } $    &   $ 16 \times 16  $    &    142        &  1.43      &    0.54       \\    
  $s_0/4$         &   $  {1 \over {2~s_0/4} } $    &   $ 32 \times 32  $    &     69        &  1.58      &    0.50       \\
\hline\hline
\end{tabular}
%%%%\end{center}
\begin{list}{}{}
\item[$^{\mathrm{a}}$] $\Delta r = dx_s = s_0$ kept fixed.
\end{list}
\end{table}
%%%%\end{center}

%%%%%\caption{Averaged intensity  $<I>$ and modulation index $m$
%%%%%for each  series computed with the fixed integration surface $S = (4 r_S)^2$
%%%%%while increasing the upper frequency $q_{max} = {1 \over {2 \Delta r}}$ of the Kolmogorov spectrum $P_{3N}$. 
%%%%%The intensities are computed every other 2.5 days (V=50 km/s). The total length 
%%%%%of the screen $N\times \Delta r$ decreases by a factor 2 and 4 in this test and subsequently 
%%%%%also the number of intensities in column 4. }
%%%%%\end{table}

As a third test,  we study the impact of 
the spatial resolution $dx_s$ used to read  the phase screen file into eq.~(5). 
Although we have just concluded that 
$\Delta r = s_0/4$ is necessary to include enough  high frequencies,
 the phases can possibly be read with a lesser resolution. Using the phase screen file constructed with
 $C_n^2 = 10^{-3}$, $z = 3.6$ kpc, $N=2^{17}$  and $\Delta r = s_0/4$, 
 we test the three cases: $dx_s = s_0$ in reading every other four phases,
  $dx_s = s_0/2$ in reading every other two phases
and  $dx_s = s_0/4$ in reading every  phase. The integration surface is fixed to $S = (4 r_S)^2$ 
in this test.
 Fig.~B.4 shows the three dynamic spectra corresponding
to these resolutions. As the Figure shows, they have the same morphology 
and their averaged intensities changed by only 6 \%. We extended
the resolution  further to  $dx_s = 2 s_0$ and found then that
the morphology of the dynamic spectrum becomes abruptly very dissimilar
(not shown in  Fig.~B.4) meaning it had not yet converged.
  We also computed the intensity  series for significant
fractions of the screen and report  $<I>$ and $m$ in Table~B.3.  
From this third test, we conclude that $dx_s = s_0$ is sufficient.

%%\medskip  
%%\centerline {[ Place Fig B4 here ]}
%%\bigskip

%%%\begin{center}
\begin{table} [h]
  \caption[]{Sensitivity of the averaged intensity  and modulation index to  $dx_s$ $^{\mathrm{a}}$}
\vspace{0.2cm} 
%%%%\begin{center}
\begin{tabular}{lcccc} \hline\hline
Resolution $dx_s$ &  Sampling           &    Nb of     &    $<I>$      &  Index $m$   \\
                 &   of dyn. sp.       &    intensities    &               &              \\\hline
 $1 s_0$          &   $8 \times 8$      &      69          &  1.57         &      0.54    \\
   $s_0/2$        &    $16 \times 16$   &      31          &  1.87         &      0.49    \\    
 $s_0/4$          &   $32 \times 32$    &       18         &  1.98        &      0.49    \\\hline\hline
\end{tabular}
%%%\end{center}
%%%% \caption{Averaged intensity  $<I>$ and modulation index $m$
%%%%for each  series computed with the fixed integration surface $S = (4 r_S)^2$
%%%% and the fixed $q_{max} = {1 \over {2 s_0/4}}$  but 
%%%% with different  spatial resolutions  $dx_s$ to read the phase screen  $\phi_K(m,n)$. }
\begin{list}{}{}
\item[$^{\mathrm{a}}$] $S=(4 r_S)^2$  and  $q_{max} = {1 \over {2 s_0/4}}$  kept fixed.
\end{list}
\end{table}
%%%%\end{center}

Finally, we  note that the resulting  modulation index $m$ ($\sim 0.5$)
is larger than the theoretical value $m_r =0.32$ derived for purely
refractive scintillation by Gupta, Rickett \& Coles 1993 ($m_r~=~1.08~({{\Delta \nu_{dc}} \over {\nu}})^{1/6}$).    This is because 
the size of the dynamic  spectra simulated  
(70 min $\times$ 8.8 MHz) is  not much larger than 
 the diffractive patches  $t_d \times \Delta \nu_{dc} = 8.8 {\rm min} 
\times 1.1$ MHz~~~($t_d~=~s_0/V  $
 and $\Delta \nu_{dc} = 1 / (2 \pi \tau_S)$). 
This size (70 min $\times$ 8.8 MHz) we used to average out diffractive scintillation is not large enough   
and so we have a  blend of
both refraction and diffraction into $m$.
We note also that the normalized intensity $<I>$ is $\sim 1.5$ rather than
unity  but this is a statistical fluctuation. For instance,  we found  $<I> =0.77 $
and $m = 0.51$ for scintillation simulated along a full $4 r_S$ wide 
track in another part of our large  phase screen.  
In Fig.~B.5, we provide an example of 11 dynamic spectra
2.5 days apart and their summed autocorrelation function. The visibility  at  $\sigma = s_0$ ($\sim 60 \%$) in 
$<E(s) E^{*}(s+\sigma)> = exp[-0.5(\sigma/s_0)^{\alpha}]$ (Rickett 1988)
is closely delineated by  the contours  $65 \%$ in the
2-D autorrelation function of Fig.~B.5. The half widths of this function 
are  $\sim 4 $ min  
and $\sim 0.5 $ MHz and are consistent  with the theoretical values 
 $t_d=8.8~{\rm min}$ and $\Delta \nu_{dc} = 1.1$MHz.

%%\medskip  
%%\centerline {[ Place Fig B5 here ]}
%%\bigskip

In Fig.~B.6, we compute the normalized 
autocorrelation function ($acf(\tau) / acf(0)$) of the time series 
of the intensity
shown in  Fig.~B.1 for the cases $S=(4 r_S)^2$. This function shows 
the slow decline,   $\sim 10 $~days at half-maximum of
 $acf(1) / acf(0)$,  expected 
because of the correlation induced by the long refractive
scales of the phase screen. This is consistent with the theoretical
refractive time scale $\tau_R = L \theta_S / V $ (Rickett 1988)
of 15 days with $L=1.8$~kpc,  $V=50$~km/s
and  the scattering angle $\theta_S = { \lambda \over {2 \pi s_0}} = 1.25~10^{-9}$ Rd, with $\lambda = 0.21$~m and $s_0 = 2.66~10^7$~m.   

%%\medskip  
%%\centerline {[ Place Fig B6 here ]}

\clearpage

\begin{figure*}
\centering
\includegraphics[width=14cm,angle=-90]{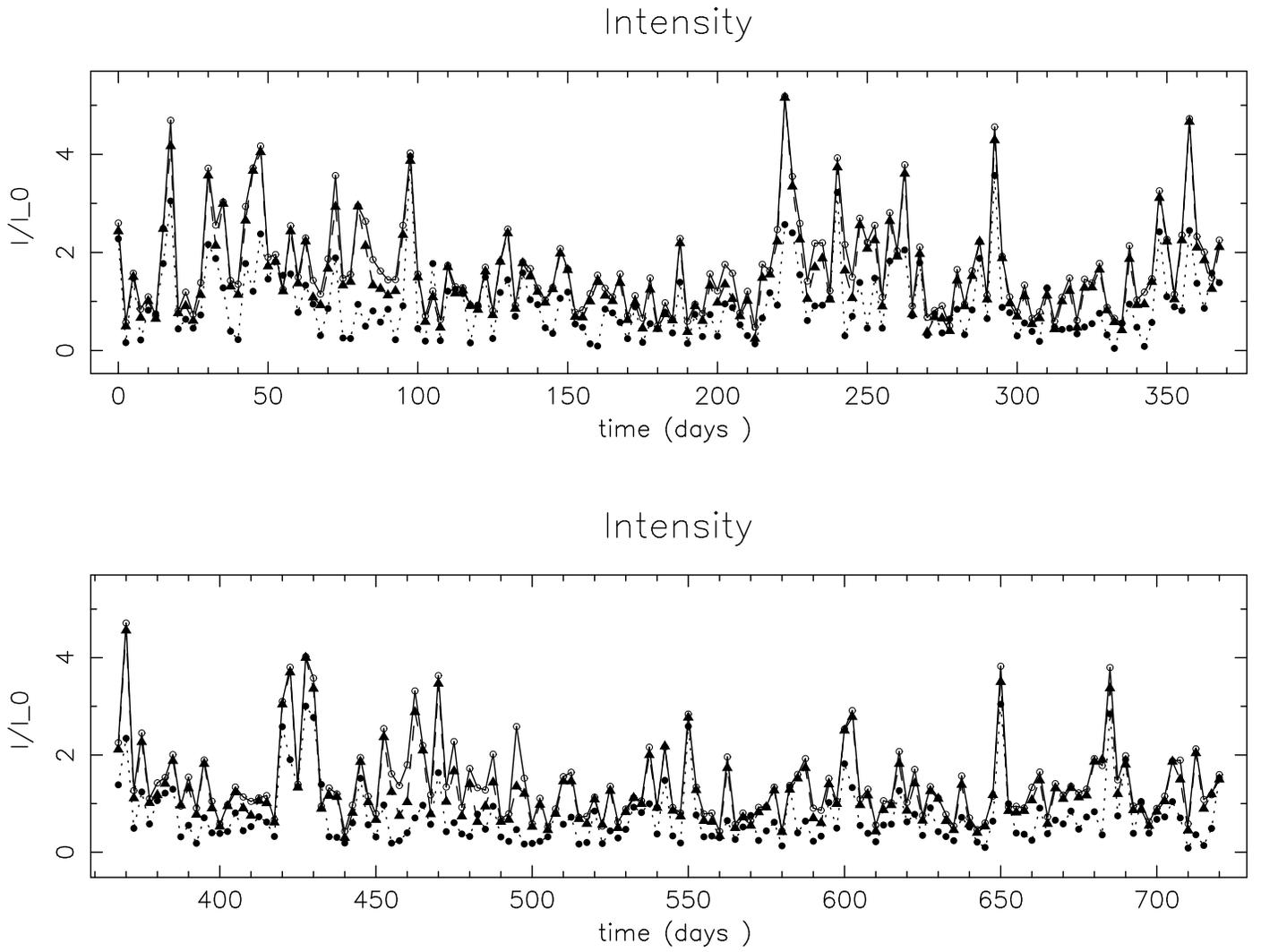}
\caption{Test on the parameter $S$ : simulations of the intensity of a pulsar at 1.41~GHz every other 2.5 days observed through
the phase screen $\phi_K(m,n)$ constructed with $C_n^2 = 10^{-3}$, $z = 3.6$ kpc, $N= 2^{17}$ and $\Delta r = s_0$.
The spatial resolution used to read the screen file  
is $dx_s = \Delta r$. The three light curves  are for the integration surfaces $S= (2 r_S)^2$ (filled circle), $S= (4 r_S)^2$ 
(filled triangle), $S = (10 r_S)^2$ (empty circle). This test demonstrates that the computation of the dynamic spectra has converged 
for the integration surface $ S = (4 r_S)^2$.}
\label{figure_mafig5/B1}
\end{figure*}

\clearpage

\begin{figure*}
\centering
\includegraphics[width=13cm, angle=-90]{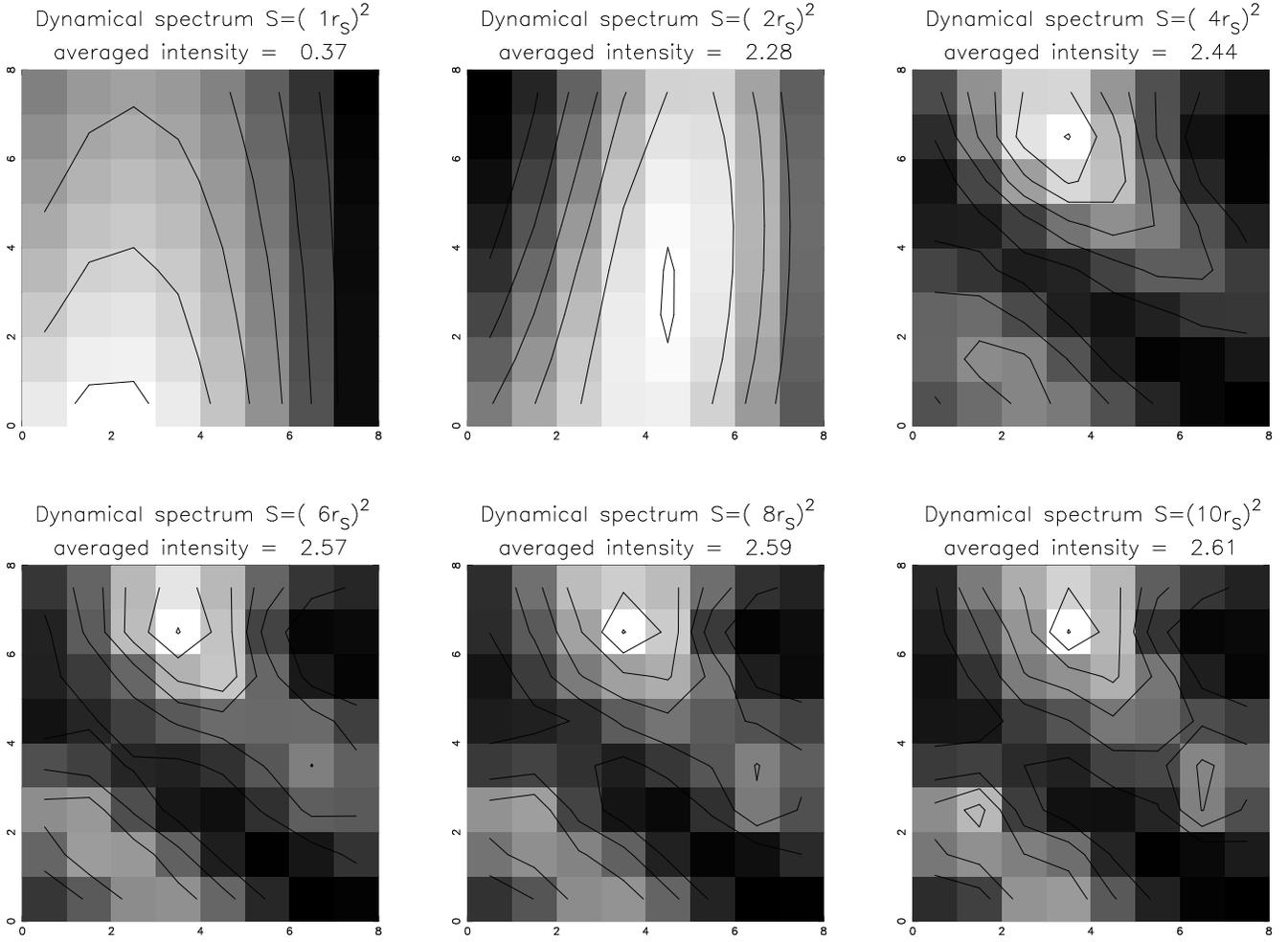}
\caption{Test on parameter $S$ : dynamic spectra computed, at 1.41~GHz,
at the same  telescope location   
 but for  different sizes $S$ of the integration surface $\mathcal{S}$
in  eq.~(1). The six cases  shown are for $S=(1 r_S)^2$, $(2 r_S)^2$, $(4  r_S)^2$, $(6 r_S)^2$, 
 $(8  r_S)^2$, $(10 r_S)^2$. 
The total size of the dynamic spectrum 
is  70 min x 8.8 MHz  sampled over $8 \times  8$ pixels,
{\sl i.e.} unity corresponds to 1.1 MHz along the frequency  axis 
(vertical) and to 8.8 minute along the time axis (horizontal)  
with the screen speed $V = 50$km/s.
The phase screen $\phi_K(m,n)$  is constructed as in Fig.~2.
This Figure shows that the 
dynamic spectrum converges from  the surface size $ S = (4 r_S)^2$.}
\label{figure_mafig6/B2}
\end{figure*}

\clearpage

\begin{figure*}
\centering
\includegraphics[width=13cm, angle=-90]{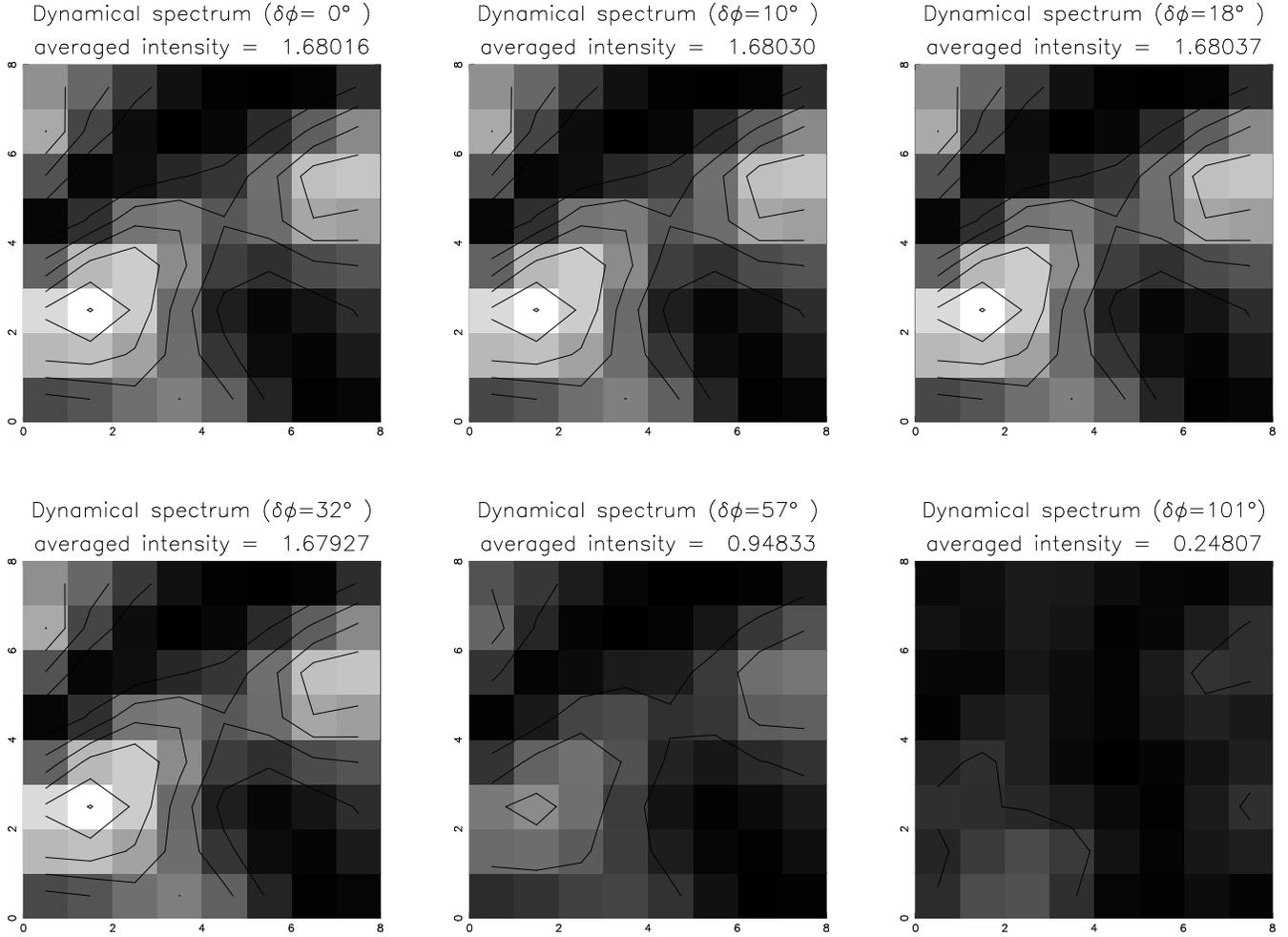}
\caption{ Test on parameter $\Delta r $:
 dynamic spectra computed at 1.41 GHz
 at the same  telescope location   
 and with the same integration surface size $S = (4 r_S)^2 $ 
 but with different grid step $\Delta r$ for 
 the  phase screen $\phi_K(m,n)$ in eq.~(4). The limits of
the Kolmogorov spectrum are directly related to $\Delta r$ and $N$ 
as $q_{min}= {1 \over {N \Delta r}}$ and $q_{max} ={1 \over {2 \Delta r}} $.   A full test 
 would require to construct effectively  several Kolmogorov 
  phase screens in decreasing  $\Delta r$ to modify $q_{max}$ and 
in increasing $N$ to keep the same $q_{min}$.
  This is very difficult to implement numerically (see text) 
and, instead,  we have
  simulated this effect in superimposing  a  corrugated surface onto  
  the original phase screen. This surface 
   is characterized by an amplitude $\delta \phi$ 
   for some spatial frequency  $q$. 
The idea is to find out which  perturbating surface  $\delta \phi$ 
  is required to modify significantly either the averaged intensity or the morphology of the
  resulting dynamic spectrum.
Our computation covered the following cases~:
   $\delta \phi = 10^{\circ}$ for $q={1 \over {2~s_0/8}}$;  
  $\delta \phi = 18^{\circ}$ for  $q={1 \over {2~s_0/4}}$;  
   $\delta \phi=32^{\circ}$ for  $q={1 \over {2~s_0/2}}$;
  $\delta \phi  = 57^{\circ}$ for  $q={1 \over { 2~s_0}}$;
  $ \delta \phi = 101^{\circ}$ for $q={1 \over {4~s_0}}$. 
  These amplitudes $\delta \phi$   are derived from 
  the phase structure function $D_{\phi}(s)$
  for the separations $s=1/8 s_0$, $s =1/4 s_0$, $s =1/2 s_0$, 
  $s =1 s_0$, $s = 2 s_0$, respectively. 
The first panel of the Figure shows  the dynamic spectrum 
   computed with the original Kolmogorov screen 
($\delta \phi =  0^{\circ}$) constructed with $C_n^2 = 10^{-3}$, $z = 3.6$ kpc,
 $N= 2^{17}$ and $\Delta r = s_0/4$.
This test shows that $q_{max} = {1 \over {2~s_0/2}}$, {\it i.e.}  $\Delta r = s_0/2$,
is necessary to include all relevant Fourier components for the computation of
dynamic spectra.  The gray scale and contours 
are the same for all panels.
Unity corresponds to 1.1 MHz along the frequency  axis 
(vertical) and to 8.8 minutes  along the time axis (horizontal)  
with the screen speed $V = 50$km/s.}
\label{figure_mafig7/B3}
\end{figure*}

\clearpage

\begin{figure*}
\centering
\includegraphics[width=6.cm, angle=-90]{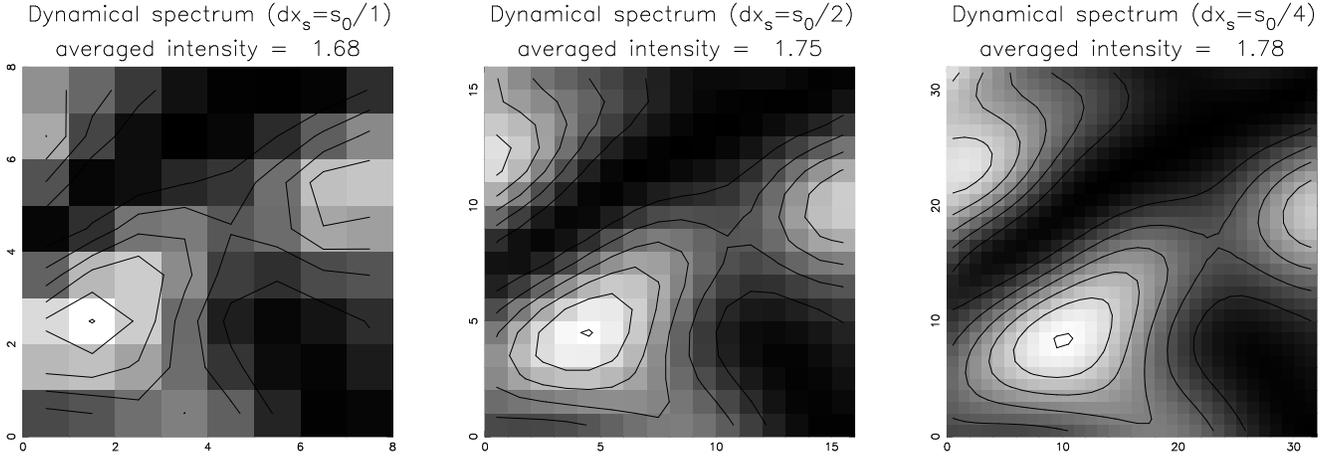}
\caption{Test on the parameter $dx_s$: dynamic spectrum computed at 1.41 GHz
at the same  telescope location and 
 with the same  integration surface $S = (4 r_S)^2 $
but with different spatial 
resolution $dx_s$ in reading the phase screen file.  
The phase screen $\phi_K(m,n)$ has been 
constructed with $C_n^2 = 10^{-3}$, $z = 3.6$ kpc,
 $N= 2^{17}$ and $\Delta r = s_0/4$. Only every other 4 pixels
are read from the phase file into eq.~(5) in the case $dx_s = s_0$;
only every other 2 pixels for $dx_s = s_0/2$;
every  pixel for $dx_s = s_0/4$ matching $\Delta r = s_0/4$ in that case.
The size of the dynamic spectrum 
is  the same for the three panels, 70 min $\times$ 8.8 MHz,
  sampled over 8 $\times$ 8, 16 $\times$ 16 
and 32 $\times$ 32 pixels, respectively, and  
unity is correspondingly  1.1 MHz, 0.55 MHz, 0.275 MHz 
along the frequency  axis 
(vertical) and  8.8 min, 4.4 min, 2.2 min along the time axis (horizontal)  
with the screen speed $V = 50$km/s.
This test shows that the spatial resolution $dx_s =  s_0$ is sufficient to make the computation of dynamic spectra and averaged intensities 
$<i({\bf x'},\lambda)>$ convergent. }
\label{figure_mafig8/B4}
\end{figure*}

\clearpage

\begin{figure*}
\centering
\includegraphics[width=10.cm,angle=-90]{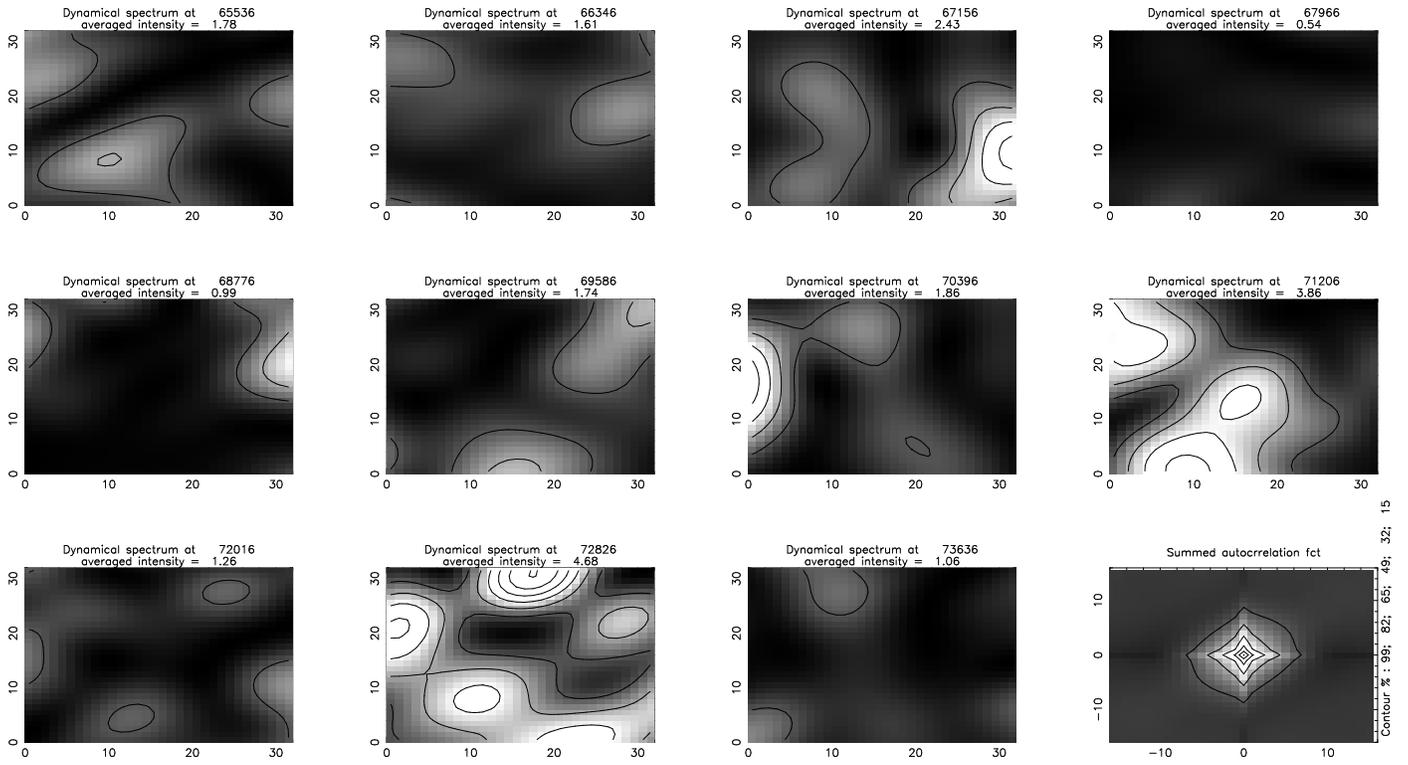}
\caption{Sample of simulated 
dynamic spectra at 1.4GHz and their summed autocorrelation 
function (last panel).
 All the dynamic spectra are plotted with the same intensity
scale, they are 1.25 days apart with the adopted screen speed $V = 50$ km/s.
They show a high variability in morphology and averaged intensity as expected
for pulsars.
The structure in these dynamic spectra is similar to speckles
 seen in images at optical wavelengths. Their summed autocorrelation
provides a mean to measure the diffactive time scale and decorrelation
bandwidth. We found they  match the theoretical values (see text).
Unity corresponds to 0.275 MHz along the frequency  axis 
(vertical) and to 2.2  minutes  along the time axis (horizontal)  
for both the dynamic spectra and the summed autocorrelation function.
 This simulation is done
with $C_n^2 = 10^{-3}$, $z = 3.6$ kpc,
$dx_s= \Delta r = s_0/4$, $N=2^{17}$ and $S= (4 r_S)^2$. }
\label{figure_mafig9/B5}
\end{figure*}

\clearpage

\begin{figure*}
\centering
\includegraphics[angle=-90.,scale=.7]{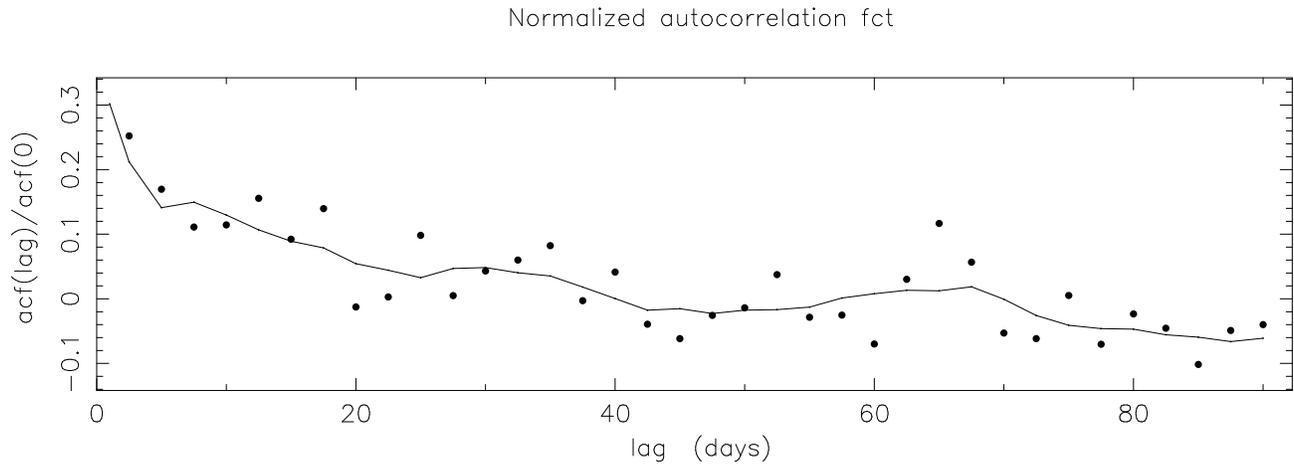}
\caption{Autocorrelation function of the  time series of the intensity 
computed with $S= (4 r_S)^2$ and shown  as filled triangles in Fig.~B.1.
The refraction time scale of $\sim 10$ days at half-maximum of this function
is consistent with  the theoretical value of 15 days.}
\label{figure_mafig10/B6}
\end{figure*}

\end{document}